\documentclass[prb,preprint,eqsecnum]{revtex4}%
\usepackage{amsmath}
\usepackage{graphicx}
\usepackage{amssymb}
\usepackage{amsfonts}%
\begin{document}
\title{Logarithmic interaction under periodic boundary conditions: Closed form
formulas for energy and forces}
\author{Sandeep Tyagi}
\email{s.tyagi@fias.uni-frankfurt.de}
\affiliation{Frankfurt Institute for Advanced Studies, J. W. Goethe Universit\"{a}t,
D-64038 Frankfurt am Main, Germany}

\begin{abstract}
A method is given to obtain closed form formulas for the energy and forces for
an aggregate of charges interacting via a logarithmic interaction under
periodic boundary conditions. The work done here is a generalization of
Glasser's results [M. L. Glasser, J. Math. Phys. \textbf{15}, 188 (1974)] and
is obtained with a different and simpler method than that by Stremler [M. A.
Stremler, J. Math. Phys. \textbf{45,} 3584 (2004)]. The simplicity of the
formulas derived here makes them extremely convenient in a computer simulation.

\end{abstract}
\maketitle

\section{Introduction}

Numerical simulations are routinely employed in the study of physical problems
that are difficult to solve analytically. Since it is not possible to simulate
realistic physical systems, containing ions of the order of Avogadro number,
one usually works with a very small system. For small systems, containing a
few hundred to a few thousand charges, boundary effects become relatively
pronounced, especially if the nature of interaction is long range. To avoid
this problem, periodic boundary conditions (PBC) are usually employed.

In many simulations, the nature of interaction is such that the potential
satisfies the Poisson equation. For example, a logarithmic interaction in two
dimensions (2D) satisfies the Poisson equation in 2D. The fact that these
potentials satisfy the Poisson equations makes it easier to treat them in the
Fourier space. Our aim in this paper is to consider one such especial case. We
obtain formulas for the energy and forces for a system of particles
interacting via the long range logarithmic potential under PBC. Related work
on the logarithmic potential could be found in the papers by
Gr\o nbeck-Jensen\cite{niels}, Liem\cite{liem} and
Tyagi\cite{tyagipre,tyagijcp}. However, all these works do not give formulas
in a closed form. Recently, Stremler\cite{stremler} generalized
Glasser's\cite{glasser} work to obtain formula for a kind of lattice sums that
have a direct bearing on the logarithmic interaction under PBC. In this paper,
we take a different approach to obtain Glasser's and Stremler's work and show
the connection of their work to the logarithmic interaction under PBC. The
derivation given here generalizes Glasser's work for the case of a rhombic
cell. The method presented is much simpler and straightforward than that given
by Stremler\cite{stremler}. The outline for the rest of the paper is as
follows. In Sec. II we present the method to obtain the closed form formulas,
and in Sec. III we discuss the results.

\section{Closed form formulas}

For two charges interacting with a pair-wise potential under PBC, the
interaction of a charge with the other one consists of three parts. The first
is the direct interaction between the charges. The second part consists of the
interaction of the first charge with all of the periodic images of the second
particle, and the last part consists of the interaction of the first charge
with all of its own periodic images. It does not matter which of the two
charges we call the first or the second charge. These three interactions when
summed over all pair of charges contained in the simulation cell form a
series. There are two important points about this series. The first is that
the series will diverge if the system is not overall charge neutral and the
second that even if the system is charge neutral the series would still be
only conditionally convergent. In the simulation of a physical system, one is
interested in the interaction energy of the charges in the basic simulation
cell with all the charges in a big box of size $R\times R.$ The box is chosen
such that the basic simulation cell is located at the center of the box. It is
clear that for a charge neutral system the result will converge to a well
defined number as $R$ tends to infinity. One way of obtaining this desired
limit is through the use of artificial background charges. The idea is to
consider that every single charge $q$, and its images come in conjunction with
a uniform distribution of charge such that the total charge, in any layer, due
to this uniformly charged layer adds up to $-q$. For a charge neutral periodic
system, imposing these uniform background charge distributions for every
charge in the system does not matter since the total uniform background charge
adds up to zero at every point of the space. However, now a charge located
within the basic simulation cell at position $\boldsymbol{r}$ may be thought
of not only interacting with a second charge located at the origin and its
periodic images, but also interacting with the neutralizing background charge
of the second particle and its images. Thus, if we have $N$ charges $q_{i}$
located at positions $\boldsymbol{r}_{i}$ in the simulation cell, then the
interaction is given by

\begin{equation}
E_{\text{total}}=\sum_{i=1}^{N-1}\sum_{j=i+1}^{N}q_{i}q_{j}G\left(
\frac{\boldsymbol{r}_{ij}}{\lambda}\right)  +\frac{1}{2}\sum_{i}q_{i}%
^{2}G_{\text{self}}, \label{first}%
\end{equation}
where $\boldsymbol{r}_{ij}=\boldsymbol{r}_{i}-\boldsymbol{r}_{j}$, and
$\lambda$ has arbitrarily been chosen as the length scale and
$G(\boldsymbol{r}/\lambda)$ satisfies the modified 2D Poisson equation%

\begin{equation}
\nabla^{2}G\left(  \frac{\left\vert \boldsymbol{r}\right\vert }{\lambda
}\right)  =-\frac{2\pi}{\lambda^{2}}\sum_{\boldsymbol{l}}\delta\left(
\frac{\left\vert \boldsymbol{r}+\boldsymbol{l}\right\vert }{\lambda}\right)
+\frac{2\pi}{l_{1}l_{2}\sin\theta}, \label{e01}%
\end{equation}
where $l_{1}$ and $l_{2}$ denote the lengths of the sides of the rhombic cell,
and $0\leq\theta<\pi$ is the angle between the two unit vectors
$\boldsymbol{e}_{1}$and $\boldsymbol{e}_{2}$ , characterizing the unit rhombic
cell with $\boldsymbol{e}_{1}.\boldsymbol{e}_{2}=\cos\theta$. The vector
$\boldsymbol{r}$ and $\boldsymbol{l}$ are given by
\begin{equation}
\boldsymbol{r}=r_{1}\boldsymbol{e}_{1}+r_{2}\boldsymbol{e}_{2}%
,\,\,\,\,\,\boldsymbol{l}=n_{1}\boldsymbol{e}_{1}+n_{2}\boldsymbol{e}_{2},
\end{equation}
where $n_{1}$ and $n_{2}$ are integers, each ranging over $-\infty$ to
$+\infty$. The second term on the right hand side of Eq. (\ref{e01}) amounts
to the presence of a neutralizing background charge mentioned above. The $G$
part represents the interaction between two different particles, while the
$G_{\text{self }}$part describes the interaction of a particle with its own
images. One notes that $G_{\text{self}}$ may be easily obtained from $G$ as%
\begin{equation}
G_{\text{self}}=\lim_{\left\vert \boldsymbol{r}\right\vert \rightarrow
0}\left[  G\left(  \frac{\boldsymbol{r}}{\lambda}\right)  +\ln\left(
\frac{\left\vert \boldsymbol{r}\right\vert }{\lambda}\right)  \right]  .
\label{a03}%
\end{equation}
The force components may be obtained from $G$ as well by taking the gradient
with respect to the position of the particle. Thus, the whole problem reduces
to obtaining $G$. The first step in this direction is to realize that the
solution of Eq. (\ref{e01}) in the Fourier space is given by
\begin{equation}
G\left(  \frac{\left\vert \boldsymbol{r}\right\vert }{\lambda}\right)
=\frac{2\pi}{l_{1}l_{2}\sin\theta}\lim_{\xi\rightarrow0}\left(  \sum
_{\boldsymbol{Q}}\frac{\exp(i\boldsymbol{Q}\cdot\boldsymbol{r})}%
{\boldsymbol{Q}^{2}+\xi^{2}}-\frac{1}{\xi^{2}}\right)  , \label{expansion}%
\end{equation}
where
\begin{equation}
\boldsymbol{Q}=n_{1}\boldsymbol{b}_{1}+n_{2}\boldsymbol{b}_{2}. \label{e02}%
\end{equation}
Here, because of the PBC, we may assume $\left\vert r_{1}\right\vert \leq
l_{1}/2$ and $\left\vert r_{2}\right\vert \leq l_{2}/2.$ The sum over
$\boldsymbol{Q}$ runs over all reciprocal lattice vectors spanned by
\begin{equation}
\boldsymbol{b}_{i}=\frac{2\pi}{l_{i}\sin^{2}\theta}(\boldsymbol{e}%
_{i}-\boldsymbol{e}_{j}\cos\theta), \label{e03}%
\end{equation}
for $(i,j)=(1,2),(2,1)$ and $n_{1}$ and $n_{2}$ are integers. From now onwards
we will assume that a final limit $\xi\rightarrow0$ is to be taken. At this
point we note that the modified Bessel function of the second kind satisfies
the relation%
\begin{equation}
\left(  \nabla^{2}+\lambda^{-2}\right)  K_{0}\left(  \frac{\left\vert
\boldsymbol{r}\right\vert }{\lambda}\right)  =-\frac{2\pi}{\lambda^{2}}%
\delta\left(  \frac{\left\vert \boldsymbol{r}\right\vert }{\lambda}\right)  .
\end{equation}
This means that the solution of
\begin{equation}
\left(  \nabla^{2}+\lambda^{-2}\right)  G_{0}\left(  \frac{\boldsymbol{r}%
}{\lambda}\right)  =-\frac{2\pi}{\lambda^{2}}\sum_{\boldsymbol{l}}%
\delta\left(  \frac{\left\vert \boldsymbol{r}\mathbf{+}\boldsymbol{l}%
\right\vert }{\lambda}\right)  \label{a01}%
\end{equation}
may be simply written as%
\begin{equation}
G_{0}\left(  \frac{\boldsymbol{r}}{\lambda}\right)  =\sum_{\boldsymbol{l}%
}K_{0}\left(  \frac{\left\vert \boldsymbol{r}\mathbf{+}\boldsymbol{l}%
\right\vert }{\lambda}\right)  . \label{g01}%
\end{equation}
On the other hand, we can obtain the solution of Eq. (\ref{a01}) directly in
the Fourier space by substituting%
\begin{equation}
G_{0}\left(  \frac{\boldsymbol{r}}{\lambda}\right)  =\sum_{\boldsymbol{Q}%
}G_{0}\left(  \boldsymbol{Q}\right)  \exp\left(  i\boldsymbol{Q}%
\mathbf{.}\boldsymbol{r}\right)  ,
\end{equation}
and using the identity%
\begin{equation}
\sum_{\boldsymbol{l}}\delta\left(  \frac{\left\vert \boldsymbol{r}%
\mathbf{+}\boldsymbol{l}\right\vert }{\lambda}\right)  =\frac{\lambda^{2}%
}{l_{1}l_{2}\sin\theta}\sum_{\boldsymbol{Q}}\exp\left(  i\boldsymbol{Q}%
\mathbf{.}\boldsymbol{r}\right)  .
\end{equation}
This results in
\begin{equation}
G_{0}\left(  \frac{\boldsymbol{r}}{\lambda}\right)  =\frac{2\pi}{l_{1}%
l_{2}\sin\theta}\sum_{\boldsymbol{Q}}\frac{\exp(i\boldsymbol{Q}.\boldsymbol{r}%
)}{\boldsymbol{Q}^{2}+\lambda^{-2}}. \label{g02}%
\end{equation}
Thus, using Eqs. (\ref{expansion}) and (\ref{g02}) we can express $G\left(
\boldsymbol{r}/\lambda\right)  $ as%
\begin{equation}
G\left(  \frac{\boldsymbol{r}}{\lambda}\right)  =\lim_{\xi\rightarrow0}\left(
G_{0}\left(  \xi\boldsymbol{r}\right)  -\frac{2\pi}{l_{1}l_{2}\sin\theta}%
\frac{1}{\xi^{2}}\right)  , \label{a05}%
\end{equation}
which by making the variables $\boldsymbol{r}$ and $\xi$ dimensionless through
scaling with length $\lambda$ and using Eq. (\ref{g01})\ can be written as%
\begin{align}
G\left(  \frac{\boldsymbol{r}}{\lambda}\right)   &  =\lim_{\xi\rightarrow
0}\left\{  G_{0}\left(  \xi\frac{\boldsymbol{r}}{\lambda}\right)  -\frac{2\pi
}{a\sin\theta}\frac{1}{\xi^{2}}\right\} \nonumber\\
&  =\lim_{\xi\rightarrow0}\left\{  \sum_{\boldsymbol{l}}K_{0}\left(  \xi
\frac{\left\vert \boldsymbol{r}\mathbf{+}\boldsymbol{l}\right\vert }{\lambda
}\right)  -\frac{2\pi}{a\sin\theta}\frac{1}{\xi^{2}}\right\}  , \label{g}%
\end{align}
where $a$ is $l_{2}l_{1}/\lambda^{2}$. From here onwards we will choose
$l_{1}$ to be our length scale thus we will set $\lambda=l_{1}$. Now writing%
\begin{align}
\frac{\left\vert \boldsymbol{r}\mathbf{+}\boldsymbol{l}\right\vert }{l_{1}}
&  =\left[  \left(  x+m\right)  ^{2}+\left(  y+n\sigma\right)  ^{2}+2\left(
x+m\right)  \left(  y+n\sigma\right)  \cos\theta\right]  ^{1/2}\\
&  =\left[  \left\{  x+m+\left(  y+n\right)  \sigma\cos\theta\right\}
^{2}+\left(  y+n\right)  ^{2}\sigma^{2}\sin^{2}\theta\right]  ^{1/2},
\end{align}
and using the identity\cite{tyagijcp}%
\begin{align}
\sum_{m=-\infty}^{\infty}K_{0}\left(  \alpha\sqrt{\left(  x+m\right)
^{2}+r^{2}}\right)   &  =\pi\frac{\exp\left(  -\alpha r\right)  }{\alpha}%
+\sum_{m=1}^{\infty}\frac{2\pi}{\sqrt{\alpha^{2}+\left(  2\pi m\right)  ^{2}}%
}\nonumber\\
&  \times\exp\left(  -r\sqrt{\alpha^{2}+\left(  2\pi m\right)  ^{2}}\right)
\cos\left(  2\pi mx\right)  ,
\end{align}
we can write%
\begin{align}
\sum_{\boldsymbol{l}}K_{0}\left(  \xi\frac{\left\vert \boldsymbol{r}%
\mathbf{+}\boldsymbol{l}\right\vert }{l_{1}}\right)   &  =\pi\sum_{n=-\infty
}^{\infty}\frac{\exp\left(  -\xi\sigma\sin\theta\left\vert y+n\right\vert
\right)  }{\xi}\nonumber\\
&  +\sum_{n=-\infty}^{\infty}\sum_{m=1}^{\infty}\frac{1}{m}\exp\left(  -2\pi
m\sigma\sin\theta\left\vert y+n\right\vert \right) \nonumber\\
&  \times\cos\left[  2\pi m\left\{  x+\left(  y+n\right)  \sigma\cos
\theta\right\}  \right]  , \label{ksum}%
\end{align}
where $x=r_{1}/l_{1}$, $y=r_{2}/l_{2}$ and we have substituted $\xi=0$ in the
second part on the rhs of Eq. (\ref{ksum}). The first sum over variable $n$ in
Eq. (\ref{ksum}) can be carried out as it is just a simple geometrical sum:%
\begin{equation}
\sum_{n=-\infty}^{\infty}\frac{\exp\left(  -\xi\sigma\sin\theta\left\vert
y+n\right\vert \right)  }{\xi}=\frac{\sigma\sin\theta}{2}\frac{\cosh\left[
\alpha\left(  1-2\left\vert y\right\vert \right)  \right]  }{\alpha
\sinh\left[  \alpha\right]  },
\end{equation}
where $\alpha=\left(  \sigma\xi\sin\theta\right)  /2$. Now we note that%
\begin{equation}
\lim_{\alpha\rightarrow0}\left(  \frac{\cosh\left[  \alpha\left(
1-2\left\vert y\right\vert \right)  \right]  }{\alpha\sinh\left[
\alpha\right]  }-\frac{1}{\alpha^{2}}\right)  =\frac{1}{3}\left(
1-6\left\vert y\right\vert +6\left\vert y\right\vert ^{2}\right)  .
\label{limit}%
\end{equation}
Thus, taking the required limit in Eq. (\ref{g}) with the help of Eq.
(\ref{limit}) we can write%
\begin{align}
G\left(  \frac{\boldsymbol{r}}{l_{1}}\right)   &  =\lim_{\xi\rightarrow
0}\left\{  \sum_{\boldsymbol{l}}K_{0}\left(  \xi\frac{\left\vert
\boldsymbol{r}\mathbf{+}\boldsymbol{l}\right\vert }{l_{1}}\right)  -\frac
{2\pi}{\sigma\sin\theta}\frac{1}{\xi^{2}}\right\} \\
&  =\frac{\pi}{6}\sigma\sin\theta\left(  1-6\left\vert y\right\vert
+6\left\vert y\right\vert ^{2}\right) \\
&  +\sum_{n=-\infty}^{\infty}\sum_{m=1}^{\infty}\frac{1}{m}\exp\left(  -2\pi
mz_{n}\right)  \cos\left(  2\pi my_{n}\right)  , \label{s1}%
\end{align}
where%
\begin{align}
z_{n}  &  =\sigma\sin\theta\left\vert y+n\right\vert ,\\
y_{n}  &  =x+\left(  y+n\right)  \sigma\cos\theta.
\end{align}
The sum over $m$ in Eq. (\ref{s1})\ can now be carried out as it just amounts
to a simple geometrical series. In fact, we can write \cite{sperb}
\begin{align}
\sum_{p=1}^{\infty}\frac{1}{p}\exp\left(  -2\pi pz\right)  \cos\left(  2\pi
py\right)   &  =-\operatorname{Re}\left[  \ln\left\{  1-\exp\left(
-\varsigma\right)  \right\}  \right] \nonumber\\
&  =-\ln\left\vert 1-\exp\left(  -\varsigma\right)  \right\vert
\text{\ \ \ \ \ \ }\ \ \ \ \ \ \ \ \ \ \text{for }z\geq0,
\end{align}
where $\varsigma=2\pi\left(  z\pm iy\right)  $ with the choice of $\pm$ sign
is up to us. Now, for all $n\geq1,$ while summing over $m$ in Eq. (\ref{s1}),
we will set $\varsigma_{n}=2\pi\left(  z_{n}+iy_{n}\right)  ~$where
\begin{align}
y_{n}  &  =\left\{  x+\left(  y+\left\vert n\right\vert \right)  \sigma
\cos\theta\right\} \\
z_{n}  &  =\sigma\sin\theta\left\vert y+n\right\vert =\sigma\sin\theta\left(
y+\left\vert n\right\vert \right)  ,
\end{align}
and for all $n\leq0$, we set $\varsigma_{n}=2\pi\left(  z_{n}-iy_{n}\right)
,$ where
\begin{align}
z_{n}  &  =\sigma\sin\theta\left\vert y+n\right\vert =\sigma\sin\theta\left(
-y+\left\vert n\right\vert \right)  ,\\
y_{n}  &  =\left\{  x+\left(  y-\left\vert n\right\vert \right)  \sigma
\cos\theta\right\}  .
\end{align}
Combining together the terms corresponding to $n$ and $-n$, we can write the
last part of Eq. (\ref{s1}) as
\begin{align}
f\left(  x,y\right)   &  =\sum_{n=-\infty}^{\infty}\sum_{m=1}^{\infty}\frac
{1}{m}\exp\left(  -2\pi mz_{n}\right)  \cos\left(  2\pi my_{n}\right)
\nonumber\\
&  =-\ln\left\vert 1-\exp\left(  -\varsigma_{0}\right)  \right\vert
-\sum_{n=1}^{\infty}\ln\left(  \left\vert 1-\exp\left(  -\varsigma
_{+n}\right)  \right\vert \left\vert 1-\exp\left(  -\varsigma_{-n}\right)
\right\vert \right) \nonumber\\
&  =-\ln\left\vert 1-\exp\left(  2\pi iz\right)  \right\vert -\sum
_{n=1}^{\infty}\ln\left\vert 1-2q^{2n}\cos\left(  2\pi z\right)
+q^{4n}\right\vert , \label{fxy}%
\end{align}
where%
\begin{equation}
z=x+y\sigma\exp\left(  i\theta\right)
\end{equation}
and%
\begin{equation}
q=\exp\left[  \pi\sigma i\exp\left(  i\theta\right)  \right]  .
\end{equation}
Using the definition of the Jacobi theta function \cite{gradshteyn}:%
\begin{equation}
\vartheta_{1}\left(  u,q\right)  =2q^{1/4}\sin u\prod_{n=1}^{\infty}\left(
1-2q^{2n}\cos2u+q^{4n}\right)  \left(  1-q^{2n}\right)  , \label{a02}%
\end{equation}
we note that%
\begin{equation}
\lim_{u\rightarrow0}\frac{\vartheta_{1}\left(  u,q\right)  }{\sin u}%
=\lim_{u\rightarrow0}\frac{\vartheta_{1}^{^{\prime}}\left(  u,q\right)  }{\cos
u}=\vartheta_{1}^{^{\prime}}\left(  0,q\right)  ,
\end{equation}
where a prime over the Jacobi theta function denotes the differentiation with
respect to the first argument. Thus, we obtain%
\begin{align}
\vartheta_{1}^{^{\prime}}\left(  0,q\right)   &  =\lim_{u\rightarrow0}%
\frac{\vartheta_{1}\left(  u,q\right)  }{\sin u}\nonumber\\
&  =2q^{1/4}\prod_{n=1}^{\infty}\left(  1-2q^{2n}+q^{4n}\right)  \left(
1-q^{2n}\right) \nonumber\\
&  =2q^{1/4}\left(  \prod_{n=1}^{\infty}\left(  1-q^{2n}\right)  \right)
^{3}. \label{a04}%
\end{align}
So, we can write Eq. (\ref{a02}) as%
\begin{equation}
\prod_{n=1}^{\infty}\left(  1-2q^{2n}\cos2u+q^{4n}\right)  =\frac
{2^{-2/3}q^{-1/6}}{\sin u}\frac{\vartheta_{1}\left(  u,q\right)  }{\left[
\vartheta_{1}^{^{\prime}}\left(  0,q\right)  \right]  ^{1/3}}. \label{s3}%
\end{equation}
Using Eqs. (\ref{fxy}) and (\ref{s3}) we can write%
\begin{equation}
f\left(  x,y\right)  =-\frac{\ln2}{3}+\pi\left\vert y\right\vert \sigma
\sin\theta-\frac{1}{6}\pi\sigma\sin\theta-\ln\left\vert \frac{\vartheta
_{1}\left(  \pi z,q\right)  }{\left[  \vartheta_{1}^{^{\prime}}\left(
0,q\right)  \right]  ^{1/3}}\right\vert . \label{s4}%
\end{equation}
Thus, using Eqs. (\ref{s1}), (\ref{fxy}) and (\ref{s4}) we finally obtain%
\begin{equation}
G\left(  \frac{\boldsymbol{r}}{l_{1}}\right)  =\pi\sigma\sin\theta\left\vert
y\right\vert ^{2}-\frac{1}{3}\ln2-\ln\left\vert \frac{\vartheta_{1}\left(  \pi
z,q\right)  }{\left[  \vartheta_{1}^{^{\prime}}\left(  0,q\right)  \right]
^{1/3}}\right\vert . \label{a06}%
\end{equation}
The result in Eq. (\ref{a06})\ is in agreement with that obtained by Stremler
\cite{stremler}. An especial case of the method gives Glasser's \cite{glasser}
results. As noted earlier by Stremler, there is a trivial mistake in the paper
by Glasser.

The self-energy of this system can now be easily obtained by employing Eq.
(\ref{a03}). First we note that as $z\rightarrow0,$ we have $\vartheta
_{1}\left(  \pi z,\tau\right)  \approx\vartheta_{1}^{^{\prime}}\left(
0,q\right)  \pi z,$ where we have used Eq. (\ref{a04}) along with the fact
that $\sin\left(  \pi z\right)  \approx\pi z.$ Also, it is easy to see that
\begin{align}
\left\vert z\right\vert  &  =\left\vert x+y\sigma\exp\left(  i\theta\right)
\right\vert \nonumber\\
&  =\left[  x^{2}+y^{2}\sigma^{2}+2xy\sigma\cos\theta\right]  ^{1/2}%
\nonumber\\
&  =\frac{\sqrt{r_{1}^{2}+r_{2}^{2}+2r_{1}r_{2}\cos\theta}}{l_{1}}.
\end{align}
The self-energy is thus given by%
\begin{align}
G_{\text{self}}  &  =\lim_{r_{1}\rightarrow0,r_{2}\rightarrow0}\left[
G\left(  \frac{\boldsymbol{r}}{l_{1}}\right)  +\ln\left(  \frac{\sqrt
{r_{1}^{2}+r_{2}^{2}+2r_{1}r_{2}\cos\theta}}{l_{1}}\right)  \right]
\nonumber\\
&  =-\frac{1}{3}\ln2-\ln\pi-\ln\left\vert \left[  \vartheta_{1}^{^{\prime}%
}\left(  0,q\right)  \right]  ^{2/3}\right\vert . \label{a07}%
\end{align}
The expressions for the force may be derived by differentiating $G\left(
\boldsymbol{r}/\lambda\right)  $ with respect to the Cartesian coordinates
$x^{\prime}$and $y^{\prime}$, which are related to the rhombic coordinates
$r_{1}$and $r_{2}$ by%
\begin{align}
r_{1}  &  =x^{\prime}-y^{\prime}\cot\theta\\
r_{2}  &  =y^{\prime}/\sin\theta.
\end{align}
Thus, we obtain%

\begin{align}
F_{x^{\prime}}\left(  \boldsymbol{r}\right)   &  =-\frac{\partial}{\partial
r_{1}}G\left(  \frac{\boldsymbol{r}}{l_{1}}\right)  \frac{\partial r_{1}%
}{\partial x^{\prime}}-\frac{\partial}{\partial r_{2}}G\left(  \frac
{\boldsymbol{r}}{l_{1}}\right)  \frac{\partial r_{2}}{\partial x^{\prime}%
}\nonumber\\
&  =\frac{\pi}{l_{1}}\operatorname{Re}\left[  \frac{\vartheta_{1}^{^{\prime}%
}\left(  \pi z,\tau\right)  }{\vartheta_{1}\left(  \pi z,\tau\right)
}\right]  ,
\end{align}
and
\begin{align}
F_{y^{\prime}}\left(  \boldsymbol{r}\right)   &  =-\frac{\partial}{\partial
r_{1}}G\left(  \frac{\boldsymbol{r}}{l_{1}}\right)  \frac{\partial r_{1}%
}{\partial y^{\prime}}-\frac{\partial}{\partial r_{2}}G\left(  \frac
{\boldsymbol{r}}{l_{1}}\right)  \frac{\partial r_{2}}{\partial y^{\prime}%
}\nonumber\\
&  =-\frac{\pi\cot\theta}{l_{1}}\operatorname{Re}\left[  \frac{\vartheta
_{1}^{^{\prime}}\left(  \pi z,\tau\right)  }{\vartheta_{1}\left(  \pi
z,\tau\right)  }\right]  -\frac{2\pi}{l_{1}}y\nonumber\\
&  +\frac{\pi\text{csch}\theta}{l_{1}}\operatorname{Re}\left[  \frac
{\vartheta_{1}^{^{\prime}}\left(  \pi z,\tau\right)  }{\vartheta_{1}\left(
\pi z,\tau\right)  }\exp\left(  i\theta\right)  \right] \nonumber\\
&  =-\frac{2\pi}{l_{1}}y-\frac{\pi}{l_{1}}\operatorname{Im}\left[
\frac{\vartheta_{1}^{^{\prime}}\left(  \pi z,\tau\right)  }{\vartheta
_{1}\left(  \pi z,\tau\right)  }\right]  .
\end{align}
We have thus obtained complete expressions for $G\left(  \boldsymbol{r}%
\mathbf{/}\lambda\right)  $, the self-energy, and the forces. These expression
for the energy, the self-energy and the forces were numerically checked and
they show perfect agreement with our earlier results \cite{tyagipre, tyagijcp}.

\section{Discussion}

The method given here has its origin in the beautiful work of Glasser
\cite{glasser} who obtained the results for logarithmic interaction in a
closed form but did so only for a orthorhombic simulation cell. However, he
also reminded his readers that the results could be easily obtained for a
rhombic cell. Although, his work was directly related to the logarithmic
interaction, it was not noticed by many researchers working on logarithmic
interaction under PBC in recent years. The reason behind why this beautiful
work went unnoticed is that it was published under the category of lattice
sums, with the word logarithmic not even mentioned in the whole paper.
Stremler \cite{stremler} has recently extended Glasser's work to include the
case where the simulation cell may be rhombic. But the derivation given by him
is tedious. In this paper we have derived direct formulas for the energy and
forces on particles interacting via the long range logarithmic interaction
under PBC. The method presented here is much simpler than that given by Stremler.

The advantages of the formulas derived here lies in their simplicity and easy
implementation in a simulation. Most functions used in the expressions for
energy and forces, such as the Jacobi Theta function, are already implemented
in most libraries of mathematical interest.


\begin{thebibliography}{9}                                                                                                %


\bibitem {niels}N. Gr{\o }nbech-Jensen, Comput. Physics Comm. \textbf{119},
115 (1999).

\bibitem {liem}S. Y. Liem and J. H. R. Clarke, Mol. Phys. \textbf{92},19 (1997).

\bibitem {tyagipre}S. Tyagi, Phys. Rev. E \textbf{70}, 066703 (2004).

\bibitem {tyagijcp}S. Tyagi, J. Chem. Phys. \textbf{122}, 014101 (2005).

\bibitem {stremler}M. A. Stremler, J. Math. Phys., \textbf{45}, 3584 (2004).

\bibitem {glasser}M. L. Glasser, J. Math. Phys., \textbf{15}, 188 (1974).

\bibitem {sperb}R. Sperb, Mol. Simulation \textbf{22}, 199 (1999).

\bibitem {gradshteyn}I. S. Gradshteyn and I. M. Ryzhik, Table of integrals
series and products Academic Press (1965).
\end{thebibliography}
\end{document}